\documentclass[twocolumn]{aastex62}

\newcommand{\hi} {{\rm H{\footnotesize I}}}
\usepackage{amsmath}
\received{xxx}
\revised{xxx}
\accepted{xxx}
\submitjournal{ApJS}

\shorttitle{Catalog of DM halo models for SPARC galaxies}
\shortauthors{Li et al.}

\begin{document}

\title{A comprehensive catalog of dark matter halo models for SPARC galaxies}

\correspondingauthor{Pengfei Li}
\email{PengfeiLi0606@gmail.com, pxl283@case.edu}

\author[0000-0002-6707-2581]{Pengfei Li}
\affil{Department of Astronomy, Case Western Reserve University, Cleveland, OH 44106, USA}

\author{Federico Lelli}
\affiliation{European Southern Observatory, Karl-Schwarschild-Strasse 2, Garching bei M\"{u}nchen, Germany}
\affiliation{School of Physics and Astronomy, Cardiff University, Queens Buildings, The Parade, Cardiff, CF24 3AA, UK}

\author{Stacy McGaugh}
\affiliation{Department of Astronomy, Case Western Reserve University, Cleveland, OH 44106, USA}

\author{James Schombert}
\affiliation{Department of Physics, University of Oregon, Eugene, OR 97403, USA}



\begin{abstract}

We present rotation curve fits to 175 late-type galaxies from the Spitzer Photometry \& Accurate Rotation Curves (SPARC) database using seven dark matter (DM) halo profiles: pseudo-isothermal (pISO), Burkert, Navarro-Frenk-White (NFW), Einasto, \citet[][DC14]{DC2014}, coreNFW, and a new semi-empirical profile named Lucky13. We marginalize over stellar mass-to-light ratio, galaxy distance, disk inclination, halo concentration and halo mass (and an additional shape parameter for Einasto) using a Markov Chain Monte Carlo method. We find that cored halo models such as the DC14 and Burkert profiles generally provide better fits to rotation curves than the cuspy NFW profile. The stellar mass-halo mass relation from abundance matching is recovered by all halo profiles once imposed as a Bayesian prior, whereas the halo mass-concentration relation is not reproduced in detail by any halo model. We provide an extensive set of figures as well as best-fit parameters in machine-readable tables to facilitate model comparison and the exploration of DM halo properties.
\end{abstract}



\section{Introduction} \label{sec:intro}

Rotation curves reveal a discrepancy between dynamically determined and optically measured masses of galaxies \citep{Rubin1978, Bosma1981, vanAlbada1985}. Together with other astrophysical evidences, this led to the introduction of dark matter. Since then, various DM halo profiles have been proposed, such as the pseudo-isotherthermal (pISO) and NFW \citep{NFW1996} profiles.

\citet{SPARC} built the Spitzer Photometry \& Accurate Rotation Curves (SPARC) database including 175 late-type galaxies with extended \hi/H$\alpha$ rotation curves and near-infrared surface photometry. This galaxy sample provides us the opportunity to make a comprehensive survey of halo models by fitting all the data in a homogeneous fashion.

A large amount of rotation curve fits can serve the purpose of exploring DM halo properties and potential correlations. For example, in \citet{Li2019}, we fit two simulation-motivated profiles, the Einasto \citep{Einasto1965} and DC14 \citep{DC2014} profiles, to the SPARC galaxies, and find that the halo scale radius and surface density of the DM halo correlate with galaxy luminosity with a similar power law, while the characteristic volume density is a constant. This finding benifits from the wide ranges in stellar mass, surface brightness and gas fraction that the SPARC galaxies span. 

In this paper, we provide rotation curve fits to 175 SPARC galaxies using seven halo models with/without $\Lambda$CDM motivated priors, depending on the availability of the priors for each profile. Summary tables and figures are organized by galaxy and by halo profile together with the best-fit parameters, so that readers can easily look up these fits for their own research. The results are made publicly available in the SPARC website.

\section{Data, models and method}

\subsection{The SPARC sample}

The SPARC database\footnote{\url{astroweb.case.edu/SPARC}} \citep{SPARC} includes 175 late-type galaxies with high-quality \hi/H$\alpha$ rotation curves and near-infrared Spitzer photometry. The \hi\ measurements allow tracing the rotation velocity ($V_{\rm obs}$) out to large radii providing strong constraints on the DM halo profiles. The Spitzer photometry has a key benefit: the stellar mass-to-light ratio has little scatter at 3.6 $\mu m$ \citep[e.g.][]{McGaughSchombert2014, Meidt2014, Schombert2019}. This effectively helps breaking the disk-halo degeneracy \citep{vanAlbada1985} when delineating the contributions of stellar disk and dark matter halo to the observed rotation curves. The mass models for the stellar disk and bulge (when present) are built by numerically solving the Poisson equation for the observed surface brightness profile at 3.6 $\mu$m. Similarly, the mass contribution of the gas is derived from the observed \hi\ surface density profile, scaled up to include Helium. The derived gravitational potentials of the baryonic components are represented by the circular velocities of test particles, tabulated as $V_{\rm disk}$, $V_{\rm bul}$, $V_{\rm gas}$ corresponding to the contributions of stellar disk, bulge and gas, respectively. For convenience, the stellar contributions in the SPARC database are tabulated using a mass-to-light ratio of unity in solar units, and need to be scaled down to more realistic values at 3.6 $\mu$m \citep{SPARC, Starkman2018}.

SPARC is a large sample by the standard of \hi\ interferometry. It includes all late-type galaxies from spirals to dwarf irregulars, and spans a large range in stellar mass (5 dex) and surface brightness ($>$ 3 dex). This makes the SPARC sample ideal for model testing and exploring the properties of DM halos. 

Galaxy distances in the SPARC database are measured via five different methods \citep[see][for details]{SPARC}: Hubble flow assuming $H_0$ = 73 km s$^{-1}$ Mpc$^{-1}$ and correcting for Virgo-centric infall, the tip magnitude of the red giant branch, the period-luminosity relation of Cepheids, membership to the Ursa major cluster of galaxies, and Type Ia supernovae. Disk inclinations are estimated kinematically. We treat distance and inclination as nuiance parameters, marginalizing over their uncertainty by imposing Gaussian priors with a standard deviation equal to their formal uncertainty.

\subsection{Dark matter halo profiles}

In this paper, we attempt to investigate all available DM profiles, including pseudo-isothermal (pISO), Burkert, NFW, Einasto, DC14, cored-NFW and a new semi-empirical profile that we call Lucky13. In general, each halo profile contains two fitting parameters: a scale radius $r_s$ and a characteristic volume density $\rho_s$. For covenience, the free parameters in our fits are the concentration $C_{200}$ and the rotation velocity $V_{200}$, which are defined as
\begin{equation}
C_{200} = r_{200}/r_s; \ V_{200}=10\ C_{200}r_sH_0,
\end{equation}
where $r_{200}$ is the radius inside of which the average halo density is 200 times the critical density of the universe. For consistency, we use these cosmologically motivated definitions also for purely empirical DM profiles, such as the pISO and Burkert models. In the following, we describe each halo model in detail.

{\bf pISO:} Rotation curves of dwarf galaxies are found to be well fit by an empirical profile with a constant-density core, the pseudo-isothermal profile \citep[see e.g.][]{Adams2014, Oh2015},
\begin{equation}
\rho_{\rm pISO} =\frac{\rho_s}{1 + (\frac{r}{r_s})^2}.
\end{equation}
The enclosed mass profile is given by 
\begin{equation}
M_{\rm pISO} = 4\pi\rho_s r_s^3\Big[x-\arctan(x)\Big],
\end{equation}
where we have introduced the dimensionless parameter $x=r/r_s$. The corresponding rotation velocity profile is
\begin{equation}
\frac{V_{\rm pISO}}{V_{200}} = \sqrt{\frac{1-\arctan(x)/x}{1-\arctan(C_{200})/C_{200}}}.
\end{equation}

{\bf Burkert:} The enclosed mass of the pISO profile quickly diverges at large radii (Eq. 3). \citet{Burkert1995} proposed a modified version of the pISO profile that diverges more slowly,
\begin{equation}
\rho_{\rm Burkert} =\frac{\rho_s}{(1 + \frac{r}{r_s})[1 + (\frac{r}{r_s})^2]},
\end{equation}
with an enclosed halo mass profile given by 
\begin{equation}
M_{\rm Burkert} = 2\pi\rho_s r_s^3\Big[\frac{1}{2}\ln(1+x^2)+\ln(1+x)-\arctan(x)\Big].
\end{equation}
Its rotation velocity is then given by
\begin{multline}
\frac{V_{\rm Burkert}}{V_{200}} = \frac{C_{200}}{x}\times \\
\sqrt{\frac{\frac{1}{2}\ln(1+x^2)+\ln(1+x)-\arctan(x)}{\frac{1}{2}\ln(1+C_{200}^2)+\ln(1+C_{200})-\arctan(C_{200})}}.
\end{multline}

{\bf NFW:} N-body DM-only simulations of structure formation predict a cuspy profile \citep{NFW1996},
\begin{equation}
\rho_{\rm NFW} =\frac{\rho_s}{(\frac{r}{r_s})[1 + (\frac{r}{r_s})]^2},
\end{equation}
which goes as $\rho\propto r^{-1}$ at small radii and $\rho\propto r^{-3}$ at large radii. Its enclosed mass profile is 
\begin{equation}
M_{\rm NFW} = 4\pi\rho_s r_s^3\Big[\ln(1+x)-\frac{x}{1+x}\Big],
\end{equation}
corresponding to the rotation velocity profile
\begin{equation}
\frac{V_{\rm NFW}}{V_{200}} = \sqrt{\frac{C_{200}}{x}\frac{\ln(1+x)-x/(1+x)}{\ln(1+C_{200})-C_{200}/(1+C_{200})}}.
\end{equation}

{\bf Einasto:} Using high-resolution DM-only simulations, \citet{Navarro2004} find that the simulated halos can be better described by the Einasto profile \citep{Einasto1965},
\begin{equation}
\rho_{\rm Einasto} = \rho_s\exp{\Big\{}-\frac{2}{\alpha_\epsilon}\Big[\Big(\frac{r}{r_s}\Big)^{\alpha_\epsilon}-1\Big]\Big\},
\end{equation}
which introduces an additional shape parameter $\alpha_\epsilon$. When $\alpha_\epsilon>0$, the profile has a finite central density. Its enclosed mass profile \citep{Mamon2005, Merritt2006} is
\begin{equation}
M_{\rm Einasto} = 4\pi\rho_sr_s^3\exp\Big(\frac{2}{\alpha_\epsilon}\Big)\Big(\frac{2}{\alpha_\epsilon}\Big)^{-\frac{3}{\alpha_\epsilon}}\frac{1}{\alpha_\epsilon}\Gamma\Big(\frac{3}{\alpha_\epsilon}, \frac{2}{\alpha_\epsilon}x^{\alpha_\epsilon}\Big),
\end{equation}
where $\Gamma(a, x) = \int^x_0t^{a-1}e^{-t}{\rm d}t$ is the incomplete Gamma function, and the velocity profile is given by
\begin{equation}
\frac{V_{\rm Einasto}}{V_{200}} = \sqrt{\frac{C_{200}}{x}\frac{\Gamma(\frac{3}{\alpha_\epsilon}, \frac{2}{\alpha_\epsilon}x^{\alpha_\epsilon})}{\Gamma(\frac{3}{\alpha_\epsilon}, \frac{2}{\alpha_\epsilon}C_{200}^{\alpha_\epsilon})}}
\end{equation}
 The shape parameter $\alpha_\epsilon$ depends on halo mass \citep{DuttonMaccio2014},
\begin{equation}
\alpha_\epsilon = 0.0095\nu^2 + 0.155,
\label{alpha}
\end{equation}
where $\log\nu = -0.11+0.146m+0.0138m^2+0.00123m^3$ and $m=\log(M_{\rm halo}/10^{12}h^{-1}M_\odot)$. Simulated DM halos present a standard deviation of 0.16 dex around the mean relation. However, in real galaxies, the final distribution of $\alpha_\epsilon$ differs significantly from this relation if we do not impose it as a Bayesian prior \citep{Li2019}. We hence include this relation as part of the $\Lambda$CDM priors (explained in Section 2.3).

{\bf DC14:} According to cosmological simulations of galaxy formation, baryonic matter accreted within the halos could exert a feedback effect on the halo and hence modify its halo profiles. \citet{DC2014} consider the baryonic feedback due to supernovae using a set of zoom-in, hydrodynamic simulations. They establish the DC14 model, whose profile is defined in terms of the model class ($\alpha$, $\beta$, $\gamma$) \citep{Hernquist1990, Zhao1996},
\begin{equation}
\rho_{\rm \alpha\beta\gamma} =\frac{\rho_s}{(\frac{r}{r_s})^\gamma[1 + (\frac{r}{r_s})^\alpha]^{(\beta-\gamma)/\alpha}},
\end{equation}
where $\beta$ and $\gamma$ are, respectively, the inner and outer slopes, and $\alpha$ describes the transition between the inner and outer regions. The values of these parameters depend on the stellar-to-halo mass ratio (SHM),
\begin{eqnarray}
\alpha &=& 2.94-\log[(10^{X+2.33})^{-1.08}+(10^{X+2.33})^{2.29}],\nonumber\\
\beta &=& 4.23+1.34X + 0.26X^2,\nonumber\\
\gamma &=& -0.06 + \log[(10^{X+2.56})^{-0.68}+10^{X+2.56}],
\label{DC14}
\end{eqnarray}
where $X=\log(M_\star/M_{\rm halo})$ is the SHM ratio in logarithmic space. Its enclosed mass profile is given by
\begin{equation}
M_{\rm DC14} = 4\pi r_s^3\rho_s\frac{1}{\alpha}[B(a, b+1, \epsilon)+B(a+1, b, \epsilon)],
\end{equation}
where $B(a, b, x)=\int^x_0t^{a-1}(1-t)^{b-1}{\rm d}t$ is the incomplete Beta function, and we define $a=(3-\gamma)/\alpha$, $b=(\beta-3)/\alpha$ and $\epsilon=\frac{(r/r_s)^\alpha}{1+(r/r_s)^\alpha}$. Thus, its velocity profile is given by
\begin{equation}
\frac{V_{\rm DC14}}{V_{200}} = \sqrt{\frac{C_{200}}{x}\frac{B(a, b+1, \epsilon)+B(a+1, b, \epsilon)}{B(a, b+1, \epsilon_c)+B(a+1, b, \epsilon_c)}}.
\end{equation}
Equation \ref{DC14} only works for the SHM ratio within ($-4.1$, $-1.3$), since this is the range where the supernovae feedback is significant and dominant. At $X<-4.1$, the energy released by supernovae is insufficient to modify the initial cuspy profile, so that an NFW profile remains. At $X>-1.3$, feedback due to active galactic nuclei might start to dominate. We hence set $X=-1.3$ as the largest acceptable value, following \citet{Katz2017}.

The fitting results for the Einasto and DC14 profiles are presented in \citet{Li2019}. For completeness and comparison, we also include those fits in this paper. 

{\bf coreNFW:} More recently, \citet{Read2016, Read2016b} investigate the evolution of isolated dwarf galaxies using high-resolution hydrodynamic simulations. They conclude that long-time evolution can transform an inner cusp into a finite central core through repeated bursts of star formation. They provide a general fitting function for the evolved DM profile in terms of the NFW profile,
\begin{equation}
M_{\rm coreNFW}(<r) = M_{\rm NFW}(<r)f^n,
\end{equation}
where $f = \tanh(\frac{r}{r_c})$ acts to cancel the central cusp. The core size $r_c$ is proportional to the stellar half-mass radius $R_{1/2}$, $r_c=\eta R_{1/2}$, where the proportional constant $\eta$ is suggested to be 1.75. There could be some galaxy-to-galaxy scatter around this value of $\eta$, but we keep it fixed to minimize the number of free parameters in the fit.

\begin{figure*}
\centering
\includegraphics[scale=0.44]{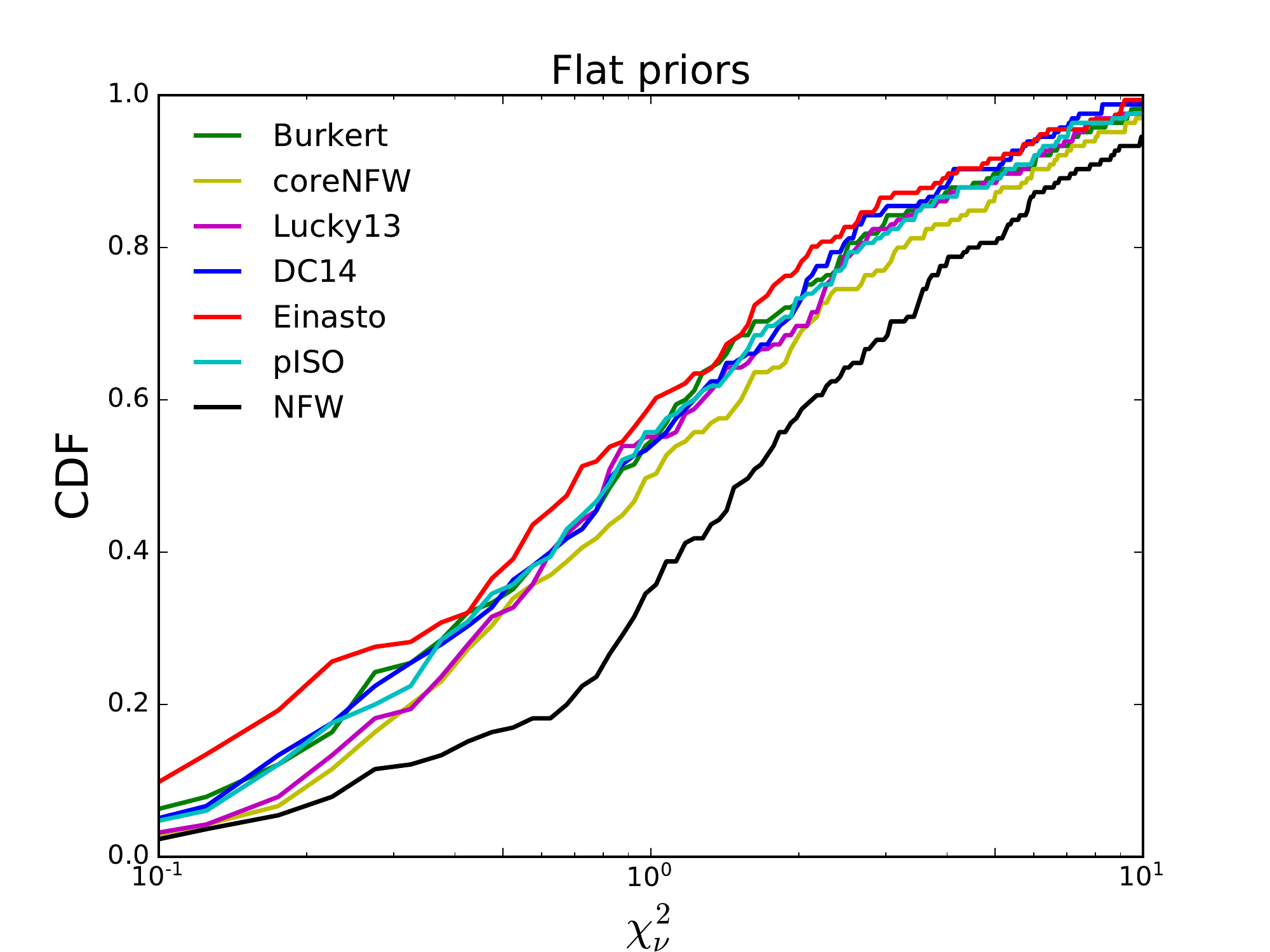}\includegraphics[scale=0.44]{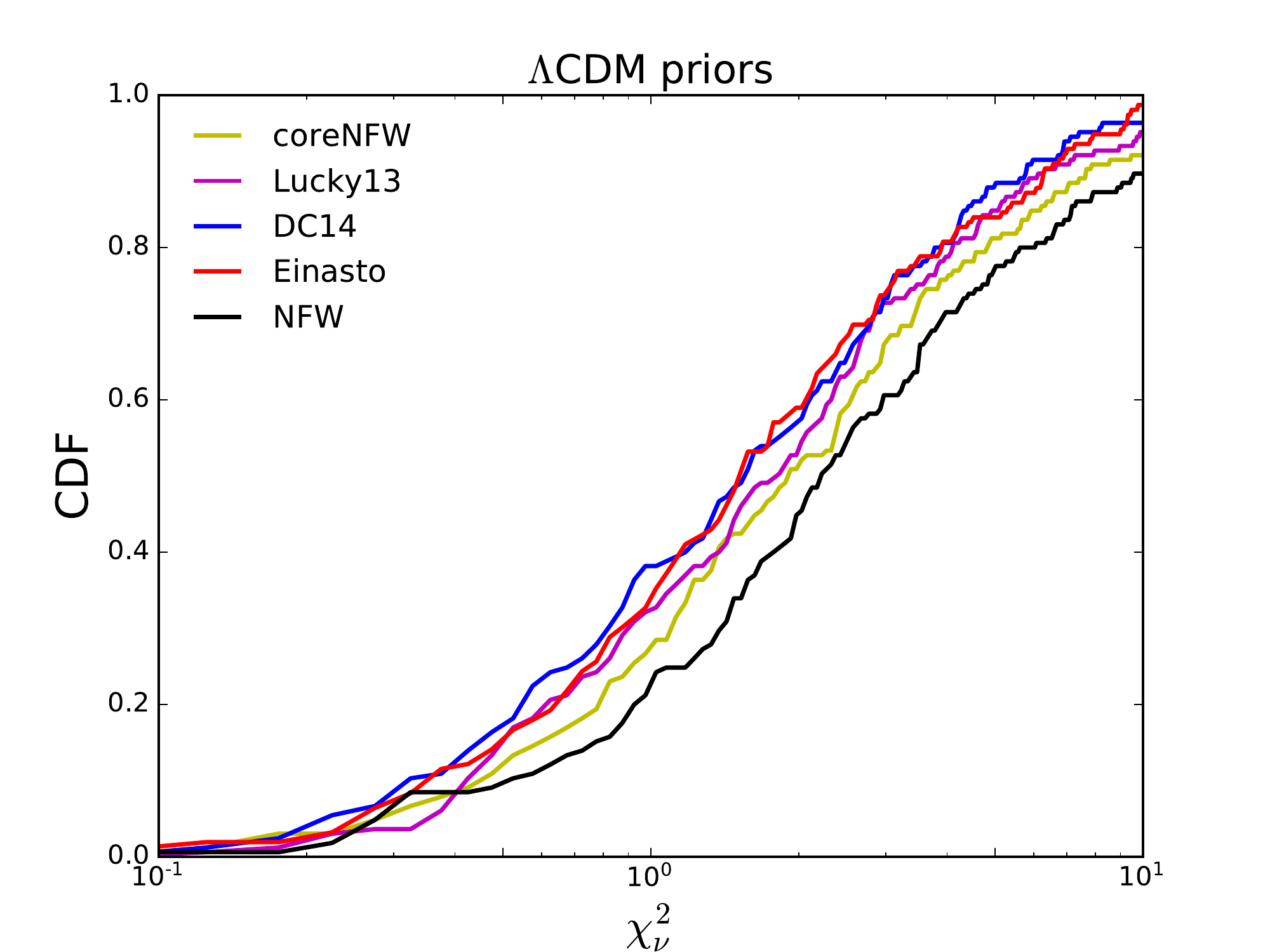}
\caption{Cumulative distributions of the reduced $\chi^2_\nu$ for seven halo profiles with flat (left) and $\Lambda$CDM priors (right).}
\label{CDFChi}
\end{figure*}

How shallow the core becomes is controlled by the evolution parameter $n$ (0 $<n<$ 1). When $n=1$, it is a complete core, while $n=0$ corresponds to a cusp. Therefore, the evolution of the halo profile is traced by the value of $n$, which is given by
\begin{equation}
n= \tanh(\kappa\frac{t_{\rm SF}}{t_{\rm dyn}}),
\end{equation}
where the so-called star-formation time $t_{\rm SF}$ is set to 14 Gyr since all SPARC galaxies are at z=0, the tuning parameter $\kappa$ is set to 0.04 as suggested by the simulations of \citet{Read2016} and the dynamic time $t_{\rm dyn}$ is defined as
\begin{equation}
t_{\rm dyn}=2\pi\sqrt{\frac{r_s^3}{GM_{\rm NFW}(r_s)}}.
\end{equation}
For the SPARC galaxies, this gives values of $n$ spanning the range 0.1 to 1.0. The resulting cored NFW (coreNFW) profile has a volume density profile given by
\begin{equation}
\rho_{\rm coreNFW} = f^n\rho_{\rm NFW} + \frac{nf^{n-1}(1-f^2)}{4\pi r^2r_c}M_{\rm NFW}.
\end{equation}

{\bf Lucky13:} We construct another cored profile from the ($\alpha$, $\beta$, $\gamma$) models by considering the specific case $\gamma=0$ to reach a finite core and $\beta=3$ to get the same decreasing rate as the NFW profile at large radii. The transition parameter is simply set as $\alpha=1$. This gives us the following profile
\begin{equation}
\rho_{\rm 130} =\frac{\rho_s}{[1 + (\frac{r}{r_s})]^3},
\end{equation}
which we call the Lucky13. Its enclosed mass profile is given by
\begin{equation}
M_{130} = 4\pi\rho_s r_s^3\Big[\ln(1+x)+\frac{2}{1+x}-\frac{1}{2(1+x)^2}-\frac{3}{2}\Big],
\end{equation}
corresponding to the velocity profile
\begin{equation}
\frac{V_{130}}{V_{200}} = \sqrt{\frac{C_{200}}{x}\frac{\ln(1+x)+\frac{2}{1+x}-\frac{1}{2(1+x)^2}-\frac{3}{2}}{\ln(1+C_{200})+\frac{2}{1+C_{200}}-\frac{1}{2(1+C_{200})^2}-\frac{3}{2}}}.
\end{equation}

\subsection{MCMC simulations}

We fit the observed rotation velocities by summing the contribution of each component,
\begin{equation}
V_{\rm tot}^2 = V_{\rm DM}^2 + \Upsilon_{\rm disk}V_{\rm disk}^2 + \Upsilon_{\rm bul}V_{\rm bul}^2 + V_{\rm gas}^2.
\end{equation}
In general, DM profiles have two free parameters $V_{200}$ and $C_{200}$ (the Einasto profile has an additional shape parameter $\alpha_\epsilon$). For the baryonic contributions, there are also three adjustable parameters: stellar mass-to-light ratio $\Upsilon_\star$, galaxy distance $D$ and disk inclination $i$. They comprise a five (six for Einasto) dimensional parameter space. To fit these halo profiles, we map the posterior distributions of these fitting parameters using the open python package $emcee$ \citep{emcee2013}. As in \citet{Li2019}, we impose lognormal priors on $\Upsilon_\star$ around their fiducial values ($\Upsilon_{\rm disk}$=0.5 and $\Upsilon_{\rm bul}$=0.7 according to \citealt{McGaugh2016PRL, OneLaw}) with a standard deviation of 0.1 dex suggested by stellar population synthesis models \citep[e.g., see][]{BelldeJong2001, Portinari2004, Meidt2014, Schombert2019}, and Gaussian priors on $D$ and $i$ around their mean values as tabulated in the SPARC database with standard deviations given by their uncertainties.

As for halo parameters, we set general loose boundaries for them: $10<V_{200}<500$ km s$^{-1}$, $0<C_{200}<1000$. Within these ranges, flat priors are imposed for all considered halo profiles. For the NFW, Einasto, DC14, coreNFW and Lucky13 profiles, we also impose the $\Lambda$CDM priors, which is comprised of the SHM relation \citep{Moster2013} and the halo mass-concentration relation \citep{Maccio2008}. The multi-epoch abundance matching determines the relation between stellar and DM halo masses,
\begin{equation}
\frac{M_\star}{M_{200}} = 2N\Big[\Big(\frac{M_{200}}{M_1}\Big)^{-\beta}+ \Big(\frac{M_{200}}{M_1}\Big)^\gamma\Big]^{-1},
\end{equation}
where $\log(M_1)$ = 11.59, $N$ = 0.0351, $\beta$ = 1.376 and $\gamma$ = 0.608. \citet{Moster2013} estimated the scatter to be $\sigma(\log\ M_\star)$ = 0.15 dex around this relation. This prior, together with the lognormal prior on stellar mass-to-light ratios, robustly breaks the disk-halo degeneracy.

\citet{Maccio2008} show that the concentration and halo mass are correlated via a power law,
\begin{equation}
\log(C_{200}) = a - b\log(M_{200}/[10^{12}h^{-1}M_\odot]),
\label{C200M200}
\end{equation}
where the coefficients $a$ and $b$ depend on cosmology and halo profile. For the NFW, coreNFW and Lucky13 profiles, we use the values from the WMAP5 cosmology corresponding to $H_0=72$ km s$^{-1}$ Mpc$^{-1}$, close to the value adopted for the SPARC database,
\begin{equation}
a = 0.830,\ \ \ b = -0.098.
\end{equation}
\citet{DC2014} show that the concentration for the DC14 profile is related to that of NFW by
\begin{equation}
C_{\rm 200, DC14} = C_{\rm 200, NFW}(1.0 + e^{0.0001[3.4(X+4.5)]}).
\end{equation}
For the Einasto profile, the coefficients as shown in \citet{Li2019} are
\begin{equation}
a = 0.977,\ \ \ b=-0.130.
\end{equation}
Equation \ref{C200M200} is the mean concentration-halo mass relation, and it has an intrinsic scatter of 0.11 dex.

\begin{figure*}[t!]
\centering
\includegraphics[scale=0.42]{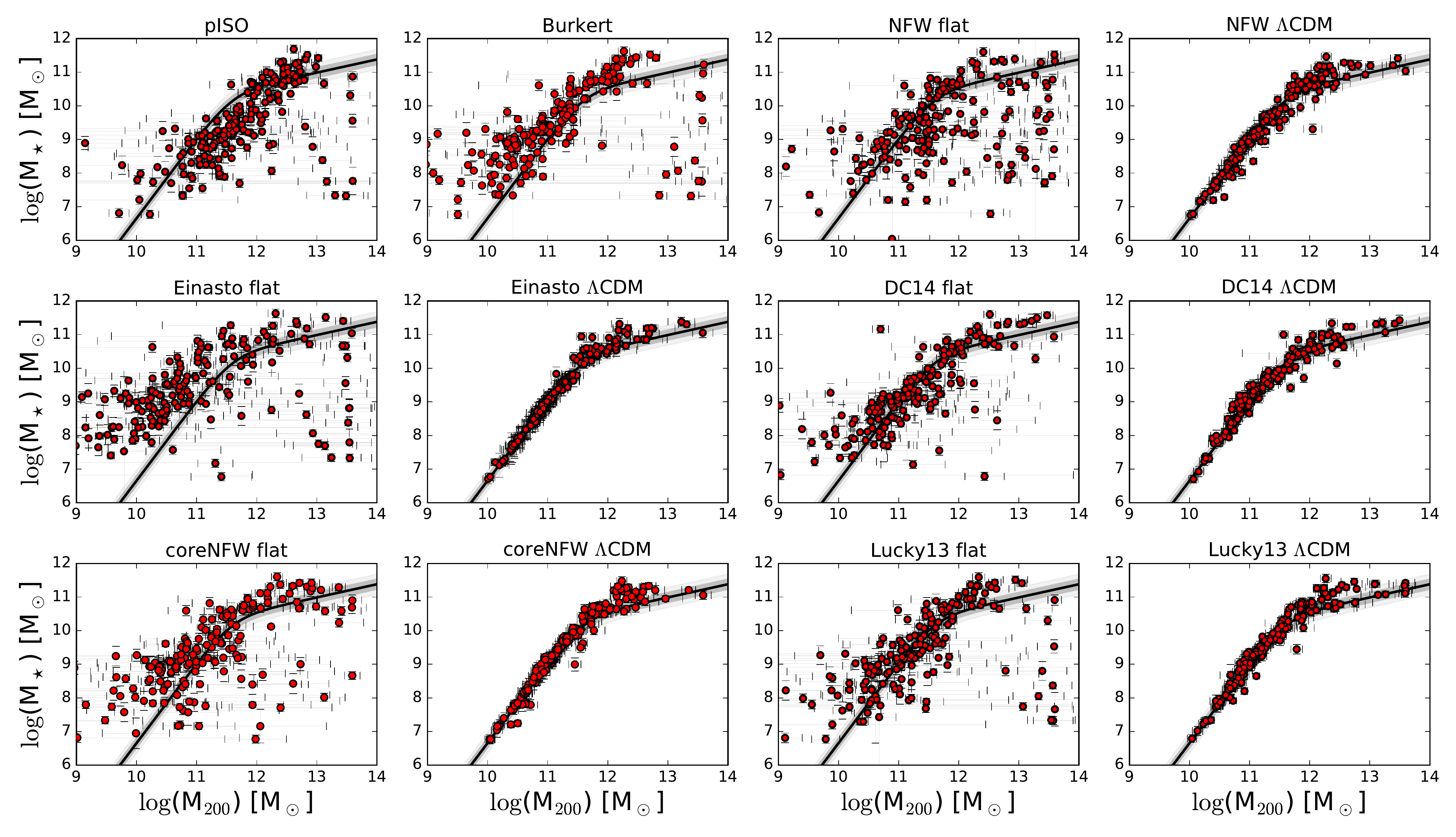}
\caption{The relations between stellar masses and DM halo masses for the seven halo models. Solid lines show the expected stellar-halo mass relation \citep{Moster2013}, which is roughly recovered when the $\Lambda$CDM priors are imposed. Dark and light shadow regions correspond to 1$\sigma$ and 2$\sigma$ standard deviations, respectively.}
\label{mstar_M200}
\end{figure*}
\begin{figure*}[t!]
\centering
\includegraphics[scale=0.42]{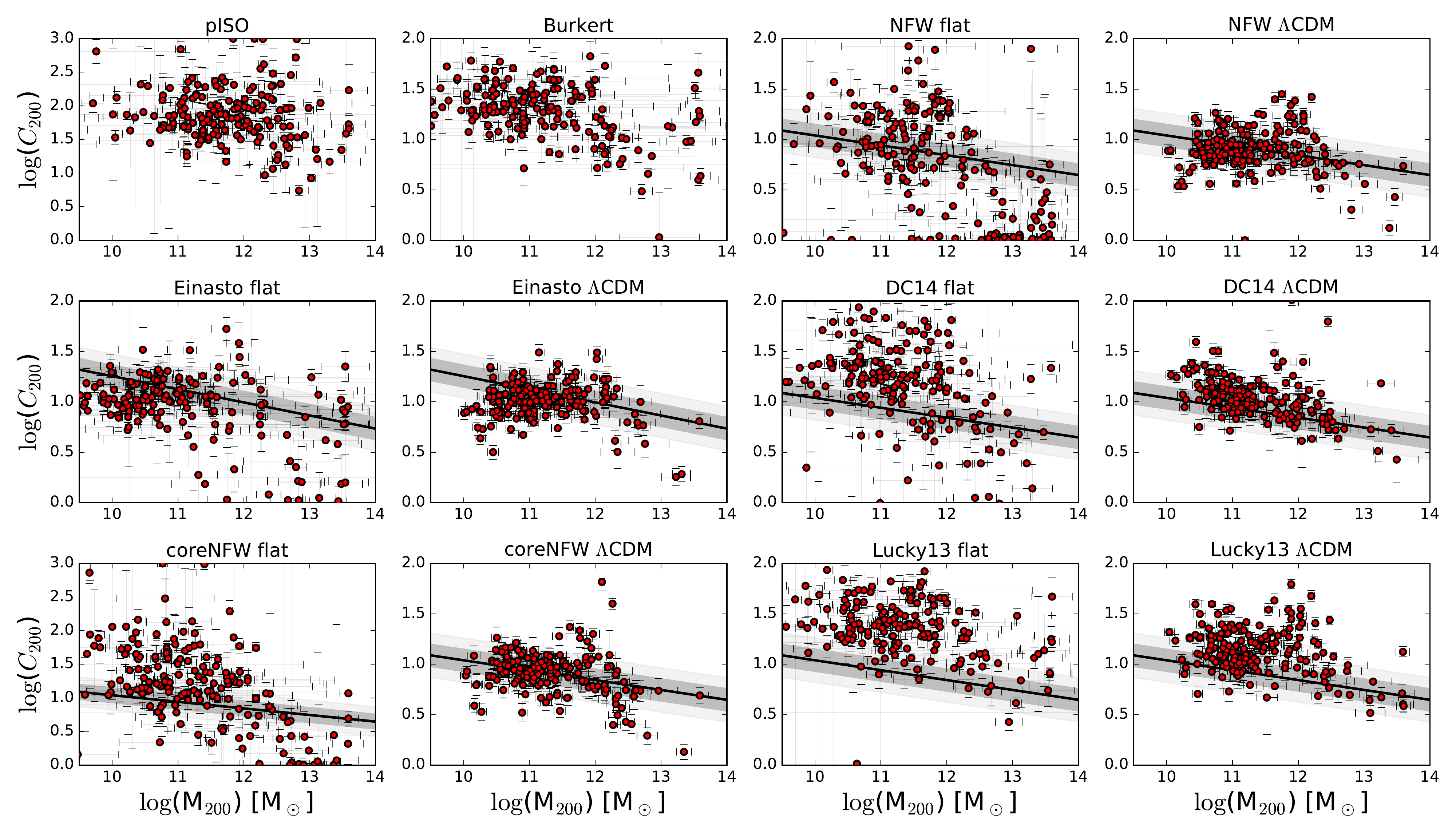}
\caption{Concentrations of the SPARC galaxies against halo masses for the seven halo models. Solid lines are the expected relations from $N$-body simulations \citep{Maccio2008}, which are model dependent and not available for the pISO and Burkert profiles. Dark and light shadow regions represent 1$\sigma$ and 2$\sigma$ standard deviations, respectively. The concentrations for the DC14 profile have been converted to that for NFW in order to compare with the imposed relation.}
\label{C200_M200}
\end{figure*}

We choose the likelihood function as $\exp(-\frac{1}{2}\chi^2)$, where $\chi^2$ is defined as
\begin{equation}
\chi^2 = \sum_R\frac{[V_{\rm obs}(R)-V_{\rm tot}(R)]^2}{(\delta V_{\rm obs})^2},
\end{equation}
where $V_{\rm obs}$ is the observed rotation velocity and $\delta V_{\rm obs}$ is the observational uncertainty. The final posterior probability is proportional to the product of the likelihood function and priors according to Bayes theorem. 

We use the standard affine-invariant ensemble sampler in $emcee$ as in \citet{Li2019}. We initialize the MCMC chains with 200 random walkers and the size of stretch-move $a=2$. We run 500 iterations for the burn-in period and then reset the sampler, before running another 2000 iterations. We check that the acceptance fractions for most galaxies are within 10\% and 70\%. There are a few galaxies with lower acceptance fractions, but their posterior distributions are well behaved. The parameter sets corresponding to the maximum probability are marked as the best-fit parameters. We estimate their uncertainties using the ``std'' output of GetDist\footnote{\url{https://getdist.readthedocs.io}}, an open Python package for analysing Monte Carlo samples.

\section{Results}

In Figure \ref{CDFChi}, we plot the cumulative distribution function (CDF) of the reduced $\chi^2$ ($\chi^2_\nu=\frac{\chi^2}{N-f}$) for all the halo profiles. Among these profiles, the Einasto profile with flat priors on halo parameters has the best fit quality since it has the largest number of fitting parameters. In general, cored profiles such as Burkert, coreNFW, DC14, Einasto, pISO, provide better rotation curve fits than the cuspy NFW profile, no matter if we impose $\Lambda$CDM priors (the combination of the stellar-to-halo mass relation and halo mass-concentration relation, as well as equation \ref{alpha} for the Einasto profile) or not. When imposing the $\Lambda$CDM priors, the fit quality decreases for all halo profiles, but the adherence to $\Lambda$CDM scaling relations drastically improve, as we now discuss.

We plot stellar versus halo masses in Figure \ref{mstar_M200}. Stellar mass shows a positive correlation with halo mass for all the halo models. When imposing flat priors, the DC14 profile presents the closest match to the SHM relation, having the smallest standard deviation of 1.01 dex. The Lucky13 profile also shows a relation that matches the expected SHM relation well except for a few outliers, resulting in a standard deviation of 1.24 dex. On the other hand, the Einasto, NFW, and coreNFW profiles display much larger scatter, having standard deviations of 1.68 dex, 1.61 dex and 1.36 dex, respectively. Finally, the Burkert and pISO profiles show mean vertical shifts of 0.51 dex and -0.44 dex with respect to the expected SHM relation, giving systematically higher and lower stellar masses. 

When we impose the $\Lambda$CDM priors, the expected SHM relation is well reproduced at low halo masses for all halo profiles. For massive galaxies, the DM halo masses are mostly smaller than the abundance-matching prediction. The extent to which this deviation is significant depends on halo models: the DC14 profile provides the best agreement, while the NFW, coreNFW, and Lucky13 profiles show larger discrepancies. The disagreement at high halo masses for the NFW profile has been pointed out by \citet{Posti2019}, who argued for a linear SHM relation for late-type galaxies. They imposed the mass-concentration relation as a prior but did not impose the SHM relation. We here confirm that there exist some discrepancies at high halo masses for the NFW profile even when we impose the SHM relation as a prior \citep[see also][]{Katz2017}.

We plot halo concentration against halo mass in Figure \ref{C200_M200}. When imposing flat priors, the pISO, Einasto, and Lucky13 profiles do not present clear trends between concentrations and halo masses, having Spearman's correlation coefficients between -0.1 and -0.3. The Burkert, NFW, DC14, and coreNFW profiles show marginal evidence for anti-correlations, having Spearman's coefficients between -0.3 and -0.5. Moreover, these putative anti-correlations seem steeper than expected from cosmology. The halo mass-concentration relation is not as well recovered as the SHM relation even if it is imposed as part of the $\Lambda$CDM priors. Remarkably, in such a case, the DC14 model is the only one to present a significant anti-correlation (Spearman's coefficient of -0.5), but the relation appears systematically shifted towards higher concentrations. The other profiles (NFW, Einasto, coreNFW, and Lucky13) have Spearman's coefficients between 0.0 and -0.2 indicative of no correlations, as evinced by the relatively flat distributions of concentrations versus halo masses.

\begin{figure*}[t]
\centering
\includegraphics[scale=0.4]{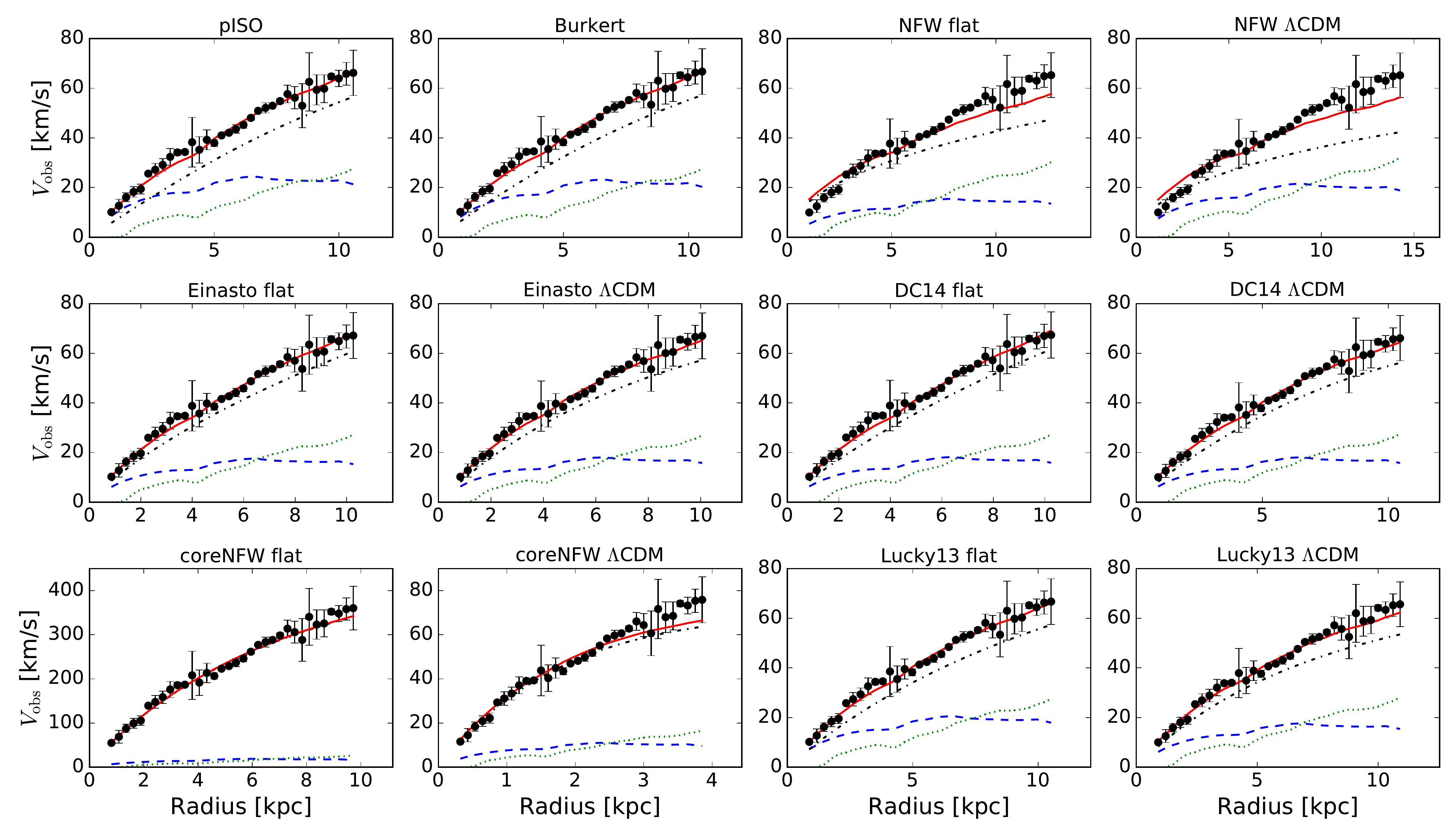}
\caption{Example galaxy: The best-fit rotation curves of the dwarf galaxy IC2574 using eight models with/without $\Lambda$CDM priors. Blue, green, purple and black lines represent disk, gas, bulge (if present) and dark matter contributions, respectively. Solid red lines are the total rotation curves, and the shadow regions reflect 1$\sigma$ (dark) and 2$\sigma$ (light) confidence levels. The complete figure set (175 images) is available in the online journal.}
\label{IC2574}
\end{figure*}

\begin{figure*}[h]
\centering
\includegraphics[scale=0.33]{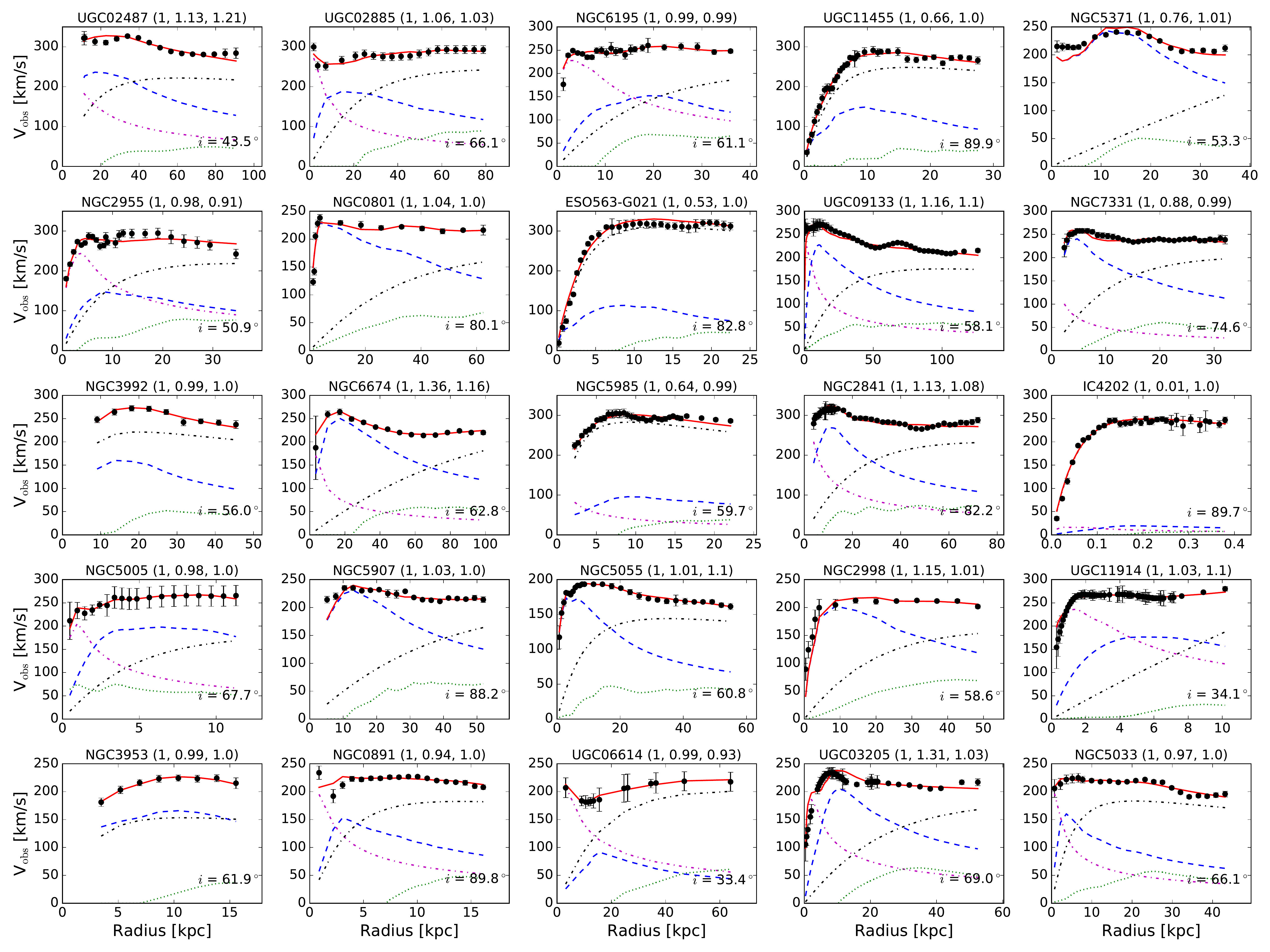}
\caption{Example model: The best rotation-curve fits of 175 SPARC galaxies using the Burkert profile. Lines are the same as those in Figure \ref{IC2574}. The three numbers in the parentheses of the subtitles are the quality flag Q \citep[for details see][]{SPARC} and the ratios of the best-fit distances and inclinations to their original values, respectively. Best-fit inclination is shown within each panel. Galaxies are ordered by decreasing quality and luminosity. The complete figure set for all the halo models (7$\times$12 images) is available in the online journal.}
\label{Burkert}
\end{figure*}

In Figure \ref{IC2574}, we show the fits for an example galaxy, IC2574, using all the models. We list the best-fit parameters in Table 1. In Figure \ref{Burkert}, we show the fits of all SPARC galaxies using the Burkert profile. Similar figures and tables are available on the SPARC website for all 175 galaxies and all seven halo profiles.

\section{Conclusion}

In this paper, we provide the community with a homogeneous catalog of DM halo parameters for 175 galaxies from the SPARC database, considering seven different halo models. Homogeneity is an important guarantee for fair comparisons of models, as \citet{Korsaga2019} find that different fitting procedures can lead to significantly different fitting results. The halo parameters are derived performing MCMC fits to \hi/H$\alpha$ rotation curves. We impose flat priors on the halo parameters, Gaussian priors on galaxy distance and disk inclination, and lognormal prior on stellar mass-to-light ratio. For five DM halo models, we also present rotation-curve fits imposing basic $\Lambda$CDM priors: the stellar mass-halo mass relation from abundance matching and the mass-concentration relation from cosmological simulations. In general, cored DM profiles provide better fits than the cuspy NFW. Moreover, while the stellar mass-halo mass relation is generally recovered by all halo models when imposed as a prior, the mass-concentration relation is not reproduced in detail by any halo model. All the fit results are publicly available on the SPARC database in the form of machine-readable tables and summary figures.

\acknowledgments
This work was supported in part by NASA ADAP grant 80NSSC19k0570.

\begin{figure*}[t]
\centering
\includegraphics[scale=0.33]{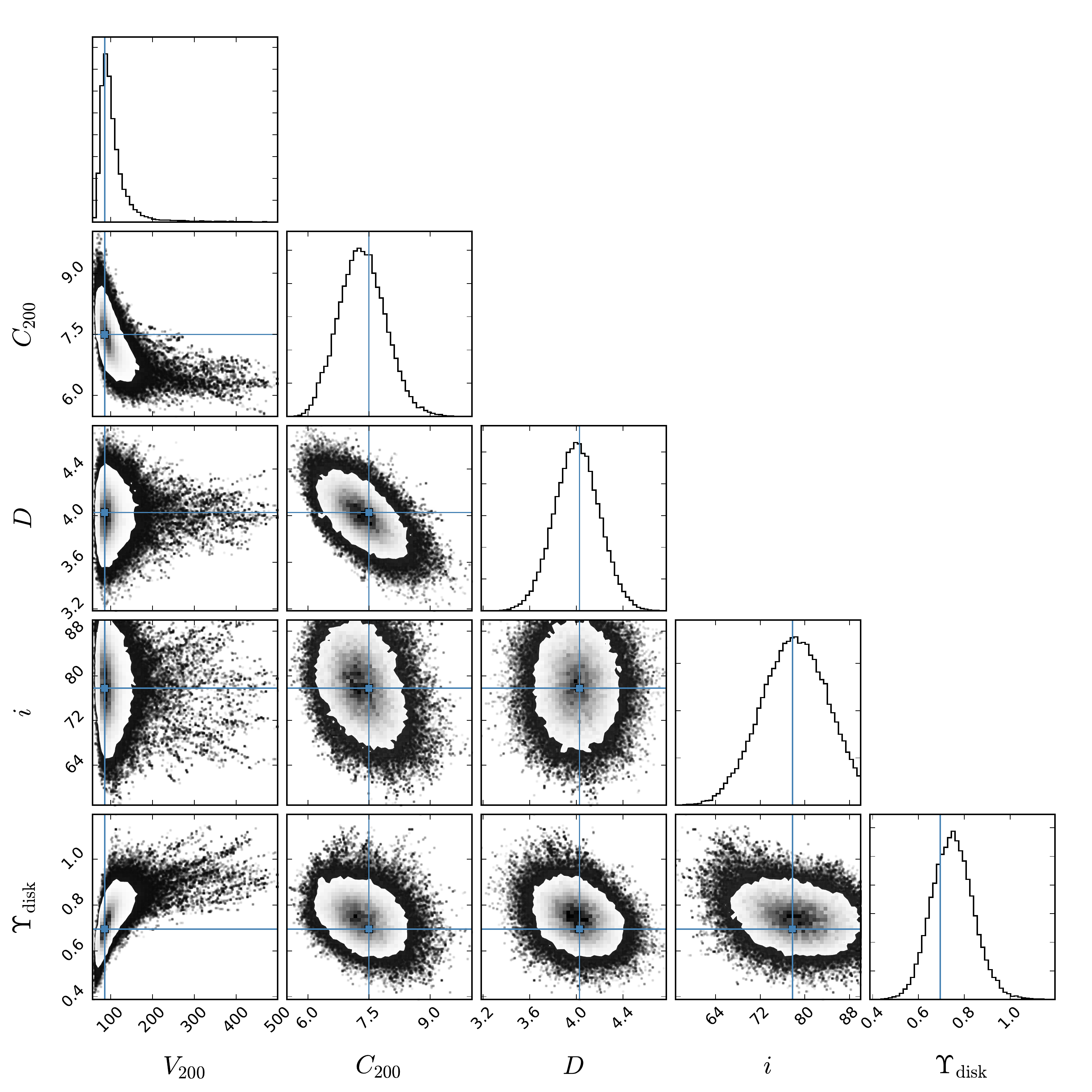}
\caption{The posterior distributions of the fitting parameters for the example galaxy IC2574 using the Burkert profile. The complete figure set for 175 SPARC galaxies using all the models (175 $\times$ 12 images) is available in the SPARC website. Also available are the corresponding Monte Carlo samples in the format required by the open Python package GetDist.}
\label{IC2574Corner}
\end{figure*}

\begin{longrotatetable}
\begin{deluxetable*}{lcccccccccccc}
\setlength{\tabcolsep}{3.2pt}
\tablecaption{The best-fit parameters for galaxy IC2574. The complete table for all 175 galaxies are available in the SPARC website.}
\tablehead{
\colhead{Model} &
\colhead{$\Upsilon_{\rm disk}$} &
\colhead{$\Upsilon_{\rm bul}$} &
\colhead{Distance} &
\colhead{Inclination} &
\colhead{$V_{200}$} &
\colhead{$C_{200}$} &
\colhead{$r_s$} &
\colhead{$\log\rho_s$} &
\colhead{$\log(M_{200})$} &
\colhead{$\alpha$} &
\colhead{$\chi^2_\nu$} \\ 
\colhead{} &
\colhead{($M_\odot/L_\odot$)} &
\colhead{($M_\odot/L_\odot$)} &
\colhead{(Mpc)} &
\colhead{(deg.)} &
\colhead{(km/s)} &
\colhead{} &
\colhead{(kpc)} &
\colhead{[$M_\odot/{\rm pc}^3$]} &
\colhead{[$M_\odot$]} &
\colhead{} &
\colhead{} 
}
\startdata
\decimals
\footnotesize
     pISO-Flat & 0.77 $\pm$ 0.09 & \nodata &  4.04 $\pm$ 0.20 & 80.0 $\pm$  5.5 & 119.39 $\pm$ 57.47 &  14.76  $\pm$  1.32 &  11.08 $\pm$   5.43 & -2.62 $\pm$  0.90 & 11.73 $\pm$ 0.63 & 0.00 $\pm$ 0.00 &  2.51\\
  Burkert-Flat & 0.69 $\pm$ 0.09 & \nodata &  4.03 $\pm$ 0.19 & 77.8 $\pm$  5.6 &  85.83 $\pm$ 44.24 &   7.49  $\pm$  0.57 &  15.69 $\pm$   8.17 & -2.52 $\pm$  0.96 & 11.30 $\pm$ 0.67 & 0.00 $\pm$ 0.00 &  2.41\\
      NFW-Flat & 0.26 $\pm$ 0.05 & \nodata &  4.83 $\pm$ 0.19 & 87.4 $\pm$  2.1 & 105.03 $\pm$  5.57 &   1.00  $\pm$  0.01 & 143.87 $\pm$   7.74 & -4.29 $\pm$  0.10 & 11.57 $\pm$ 0.07 & 0.00 $\pm$ 0.00 & 36.55\\
      NFW-LCDM & 0.44 $\pm$ 0.05 & \nodata &  5.45 $\pm$ 0.16 & 88.4 $\pm$  1.8 &  78.56 $\pm$  2.99 &   1.00  $\pm$  0.01 & 107.59 $\pm$   4.36 & -4.29 $\pm$  0.07 & 11.19 $\pm$ 0.05 & 0.00 $\pm$ 0.00 & 36.30\\
  Einasto-Flat & 0.41 $\pm$ 0.10 & \nodata &  3.92 $\pm$ 0.20 & 76.2 $\pm$  6.4 & 275.92 $\pm$ 98.58 &   1.64  $\pm$  0.81 & 230.07 $\pm$ 140.20 & -4.49 $\pm$  0.95 & 12.83 $\pm$ 0.47 & 0.33 $\pm$ 0.10 & 36.30\\
  Einasto-LCDM & 0.44 $\pm$ 0.09 & \nodata &  3.84 $\pm$ 0.20 & 76.8 $\pm$  5.9 &  63.18 $\pm$  3.50 &   4.71  $\pm$  0.25 &  18.39 $\pm$   1.42 & -3.33 $\pm$  0.13 & 10.90 $\pm$ 0.07 & 0.76 $\pm$ 0.06 & 36.30\\
     DC14-Flat & 0.44 $\pm$ 0.10 & \nodata &  3.92 $\pm$ 0.20 & 75.4 $\pm$  6.4 & 123.50 $\pm$ 11.53 &   7.68  $\pm$  0.79 &  22.03 $\pm$   3.05 & -2.83 $\pm$  0.18 & 11.78 $\pm$ 0.12 & 0.00 $\pm$ 0.00 &  2.17\\
     DC14-LCDM & 0.42 $\pm$ 0.05 & \nodata &  4.00 $\pm$ 0.19 & 80.8 $\pm$  4.8 &  67.56 $\pm$  3.12 &   9.53  $\pm$  0.59 &   9.72 $\pm$   0.75 & -2.68 $\pm$  0.10 & 10.99 $\pm$ 0.06 & 0.00 $\pm$ 0.00 &  2.43\\
  coreNFW-Flat & 0.50 $\pm$ 0.12 & \nodata &  3.71 $\pm$ 0.20 & 10.4 $\pm$  0.9 & 495.74 $\pm$ 49.16 &  11.69  $\pm$  0.62 &  58.11 $\pm$   6.54 & -2.01 $\pm$  0.20 & 13.59 $\pm$ 0.13 & 0.00 $\pm$ 0.00 &  3.08\\
  coreNFW-LCDM & 0.43 $\pm$ 0.09 & \nodata &  1.47 $\pm$ 0.20 & 59.3 $\pm$  8.8 &  54.70 $\pm$  4.82 &  14.65  $\pm$  0.80 &   5.12 $\pm$   0.53 & -1.77 $\pm$  0.18 & 10.72 $\pm$ 0.11 & 0.00 $\pm$ 0.00 &  6.22\\
  Lucky13-Flat & 0.54 $\pm$ 0.11 & \nodata &  4.02 $\pm$ 0.20 & 77.9 $\pm$  5.7 & 105.99 $\pm$ 32.05 &   7.09  $\pm$  0.73 &  20.49 $\pm$   6.54 & -2.37 $\pm$  0.57 & 11.58 $\pm$ 0.39 & 0.00 $\pm$ 0.00 &  2.27\\
  Lucky13-LCDM & 0.39 $\pm$ 0.06 & \nodata &  4.17 $\pm$ 0.19 & 83.7 $\pm$  3.9 &  74.04 $\pm$  3.07 &   7.99  $\pm$  0.47 &  12.70 $\pm$   0.92 & -2.26 $\pm$  0.11 & 11.11 $\pm$ 0.05 & 0.00 $\pm$ 0.00 &  2.70\\
\enddata
\tablecomments{Table 1 is published in its entirety in the machine-readable format. A portion is shown here for guidance regarding its form and content.}
\end{deluxetable*}
\end{longrotatetable}

\bibliographystyle{aasjournal}
\bibliography{PLi}

\end{document}